\documentclass[%
 reprint,
 amsmath,amssymb,
 aps,
]{revtex4-1}

\usepackage{graphicx}% Include figure files
\usepackage{dcolumn}% Align table columns on decimal point
\usepackage{bm}% bold math
\usepackage{color}% bold math

\usepackage{amsmath}	% Advanced maths commands
\usepackage{amssymb}	% Extra maths symbols

\def\be{\begin{equation}}
\def\ee{\end{equation}}

\begin{document}
\title{\bf{Crust effects and the cooling relaxation time in highly magnetized neutron stars }} 
\author{B. Franzon $^{1}$}
 \email{franzon@fias.uni‐frankfurt.de}
\author{Rodrigo Negreiros$^{2}$}
\email{rnegreiros@id.uff.br}
\author{S. Schramm$^{1}$}
 \email{schramm@fias.uni‐frankfurt.de}

\affiliation{$^{1}$Frankfurt Institute for Advanced Studies, Ruth-Moufang-1 60438 Frankfurt am Main, Germany}
\affiliation{$^{2}$Instituto de F\'isica, Universidade Federal Fluminense, Av. Gal. Milton Tavares S/N, Niteroi, Brazil}

\begin{abstract}
We study the effects of high magnetic fields on the structure and on the geometry of the crust in neutron stars.  We find that the crust geometry is substantially modified by the magnetic field inside the star. We build stationary and axis-symmetric magnetized stellar  models  by using well-known equations of state to describe the neutron star crust, namely the Skyrme model (Sky)  for the inner crust and the Baym, Pethick, and Sutherland (BPS) equation of state for the outer crust. We show that the magnetic field has a dual role, contributing to the crust deformation via the electromagnetic interaction (manifested in this case as the Lorentz force) and by contributing to curvature due to the energy stored in it. We also study a direct consequence of the crust deformation due to the magnetic field: the thermal relaxation time. This quantity, which is of great importance to the thermal evolution of neutron stars is sensitive to the crust properties and, as such, we show that it may be strongly affected by the magnetic field.  
\end{abstract}
%\begin{keywords}
%equation of state -- magnetic fields -- neutron stars
%\end{keywords}

%%%%%%%%%%%%%%%%%%%%%%%%%%%%%%%%%%%%%%%%%%%%%%%%%%

%%%%%%%%%%%%%%%%% BODY OF PAPER %%%%%%%%%%%%%%%%%%

\maketitle

\section{Introduction}
The density of matter varies enormously in neutron stars, from about the density of iron (7.86$\,\rm{g/cm^{3}}$) on the the stellar surface to values higher than the nuclear saturation density at the stellar the core \citep{weber1999pulsars,shapiro2008black}. In the absence of an exact theory  of  superdense  matter,  different  theoretical  models  predict different  equations  of  state (EoS) and  compositions to describe neutron stars (NS). Furthermore, the structure of NS's can be divided into the surface, composed of ions and non-relativistic electron; the outer crust, where the  ions form a solid Coulomb lattice at densities lower than the neutron drip density $n_{drip}\sim4.3\times 10^{11}\,\rm{g/cm^{3}}$; the inner crust region beyond neutron drip density, where neutrons leak out of nuclei up to densities $\sim 10^{14}\,\rm{g/cm^{3}}$ and, finally,  the core region, typically composed of electrons, protons and neutrons forming a relativistic fluid. It is also in the core that exotic degrees of freedom as hyperons \citep{glendenning1987hyperons, bednarek2012hyperons, vidana2013hyperons, gomes2015many}, quark matter \citep{Franzon:2012in, weber1999quark, baldo2004quark, alford2007astrophysics, Franzon:2016urz} and superconducting phases might appear inside these objects \citep{baldo2003neutron, kaplan2002novel, lugones2003high}.

%Approximately 1 km in thickness, Crystalline lattice of ions with elastic properties, inverse $\beta$-decay causes nuclei to become neutron rich, divides into outer and inner crust; inner crust is the region beyond neutron drip density, where neutrons leak out of nuclei, strong magnetic fields can produce distortions in a neutron star and influence the atmosphere region in particular.\\

With approximately 1 km in thickness, the crust region of neutron stars has an equation of state relatively well-known, see \cite{chamel2008physics, lorenz1993neutron, lattimer2001neutron}. In general, the composition, the structure and the equation of state of the outer crust are determined by finding the ground state of cold ionic matter. In other words, this corresponds to minimizing the Gibbs energy per nucleon at a given pressure. In this case, one nucleus occupies one neutral unit Winger-Seitz cell, which together with the nucleus and the electrons, contributes also to the total energy and pressure of the system. Here, we describe the ground state of matter in the outer crust of neutron stars by using the classical work formulated by Baym, Pethick and Sutherland \citep{baym1971ground}.

%EOS in the outer crust will be described , for example, can be  reliably determined by comparison with nuclear experiments. 

The inner crust of neutron stars begins when neutrons start to drip out of the nuclei at densities about $n_{drip}$. From this value to densities at the crust-core transition point, one has very neutron rich nuclei immersed in a gas of neutrons.  In this case, the equation of state is usually obtained with many-body techniques as Hartree-Fock (HF),  Thomas Fermi (TF)  approximation, and the Compressible Liquid Drop Model (CLDM). In this context, we follow the prescription of  \cite{douchin2001unified} to describe the  structure and composition of the inner neutron-star. The authors in \cite{douchin2001unified}  calculated the ground state of matter within the CLDM with SLy effective nucleon-nucleon interaction. It is worth to mention that other equations of state for this regime can be found in \cite{negele1973neutron, shen2002complete}. However, the choice of a particular EoS does not alter our conclusions. In addition, it is generally accepted that a pasta phase may appear in the crust-core transition \citep{grill2014equation,watanabe2000thermodynamic,ravenhall1983structure}. Although the properties of the inner crust is modified in the presence of pasta phases, we do not take them into account in this work.

Due to the its low density regime, the crust has just a small contribution to the total mass in neutron stars \citep{haensel2001neutron}. Notwithstanding, the crust region is crucial not only for determining the stellar radius, which is of major present importance due to large uncertainties in the measurements of radii in NS's, but also the crust plays a crucial role in neutron star evolution, its dynamics and observation. For example, the crust is related to phenomena as glitches \citep{ruderman1998neutron}, braking index \citep{cheng2002phase}, torsion modes \citep{levin2011excitation,sotani2007torsional,hansen1980torsional}, magnetic field evolution \citep{pons2007magnetic,aguilera2008impact,cumming2004magnetic}, thermal relaxation \citep{gnedin2001thermal} and cooling of neutron stars \citep{negreiros2012thermal,heinke2010direct,page2009neutrino}. 

Certain classes of neutron stars are associated with very strong magnetic fields. According to observation of Soft Gamma-ray Repeaters (SGR) and Anomalous X-ray Pulsars (AXP), such stars show surface magnetic fields up to $10^{15}$ G \citep{vasisht1997discovery,kouveliotou1998x}. These strong magnetic fields might be generated by dynamo processes in newly born neutron stars \citep{Thompson:1993hn},  although the exact origin of such high magnetic fields is still the subject of much debate. Moreover, according to the virial theorem, the magnetic field can reach values of $\sim 10^{18}$ G in the stellar core. According to \cite{lai1991cold,chakrabarty1997dense}, strong magnetic fields modify the equilibrium nuclear composition and the equation of state in neutron stars. However, as already shown in \cite{Franzon:2015sya,Chatterjee:2014qsa}, the global properties of compact  stars, as the mass and the radius, do not chance significantly with the inclusion of  magnetic field effects in the equation of state of the dense matter. On the other hand, it was shown in \cite{Franzon:2015sya, Franzon:2016iai} that the particle degrees of freedom at the core of stars change drastically with the inclusion of magnetic fields. Similarly, modifications of the crust properties and composition induced by magnetic fields can be seen in \cite{nandi2011neutron}. 

Strong magnetic fields  are also known to change considerably the structure of neutron stars. The authors in \cite{Chatterjee:2014qsa,Bocquet:1995je,cardall2001effects,Franzon:2015sya,mallick2014deformation}  evaluated magnetized models of stars endowed with strong poloidal magnetic fields. In this case, the Lorentz force induced by the magnetic field makes stars  more massive and they become oblate with respect to the symmetry axis. Moreover, effects of  toroidal magnetic fields were addressed  in \cite{frieben2012equilibrium,kiuchi2008equilibrium}. In this case,  the magnetized stars become more prolate with respect to the non-magnetized case. Nonetheless, these works did not address the effects of strong magnetic fields on the global properties of the neutron star crust. 

In this work, we construct equilibrium configuration of magnetized stellar models by using the same approach as in \cite{Bocquet:1995je,Bonazzola:1993zz}. We make use of spherical polar coordinates $(r, \theta, \phi)$ with origin at the stellar center and the pole located along the axis of symmetry. We focus on the size  and geometry of the crust of highly deformed strongly magnetized neutron stars. In other words, we consider the different effects of the Lorentz force according to the angle and radius distribution inside the star. 

The Lorentz force is related to the macroscopic currents that create the magnetic field, acting on the matter which can be pushed outward or inward. In the first case, we have the standard and expected effect of the Lorentz force, which acts against gravity, pushes the matter off-center making the star bigger on the equatorial plane and smaller at the pole. However, as we will see, the Lorentz force reverses direction inside the star, acting inward in the outer layers of the neutron star. It is important to notice that in addition to the just described Lorentz force, the magnetic field will also contribute to the curvature of space-time via the energy it stores. Note that, once the spherical symmetry is broken in highly magnetized neutron stars, the crust thickness depends both on the coordinate radius $r$ and on the angular direction $\theta$. 

The plan of the paper is as follows. In Section II we
give a general overview of the Einstein-Maxwell-Equations that are required to be solved numerically.  In Section III, we present our results for the crust thickness in strongly magnetized stars.  Sec. IV contains our results for the thermal relaxation time  of the stars discussed in Sec. III. Our final remarks and conclusions can be found in Sec. IV. 

\section{Stellar models with axisymmetric magnetic field}
In this work, we construct models of stationary highly magnetized neutron stars.  Details of the Einstein-Maxwell equations, numerical procedure and tests can be found in \cite{Bonazzola:1993zz,Bocquet:1995je}. We show here only the key equations that are solved numerically for the sake of completeness and better understanding for the reader. 

Equilibrium stellar configuration are obtained in general relativity by solving the Einstein equations:
\be
R_{\mu\nu} -\frac{1}{2}\mathrm{g}_{\mu\nu} R = kT_{\mu\nu},
\label{einstein}
\ee
with $R={g}_{\mu\nu}R^{\mu\nu}$, being $R_{\mu\nu}$ the Ricci tensor, ${g}_{\mu\nu}$ the metric tensor, $k$ a constant and $T_{\mu\nu}$ the energy-momentum tensor of the system.
%  may be written in this case  the sum of that of a perfect fluid with the electromagnetism stress tensor: 
%\be
%T_{\alpha\beta} = (e+p)u_{\alpha}u_{\beta} + pg_{\alpha\beta} + \frac{1}{\mu_{0}} \left( F_{\alpha \mu} F^{\mu}_{\beta} - \frac{1}{4} F_{\mu\nu} F^{\mu\nu} \mathrm{g}_{\alpha\beta} \right),
%\label{emt}
%\ee
%with  $F_{\alpha\mu}$ being the antisymmetric Faraday tensor defined as $F_{\alpha\mu} = \partial_{\alpha} A_{\mu} - \partial_{\mu} A_{\alpha} $, where $A_{\mu}$ is the electromagnetic four-potential. 
As we will be dealing with macroscopic structure of neutron stars endowed with magnetic fields, the energy-momentum tensor of the system is given by
\be 
T_{\mu\nu} = \left( \mathcal{E} + P \right) u_{\mu}u_{\nu} + P\, g_{\mu\nu}  + \frac{1}{ \mu_{0}} \left( F_{\mu \alpha} F^{\alpha}_{\nu} - \frac{\mathrm{g}_{\mu\nu}}{4} F_{\alpha\beta} F^{\alpha\beta}  \right),
\label{emt}
\ee
where the first term in Eq. \eqref{emt} is the perfect fluid contribution, with the matter energy density  $\mathcal{E}$, the isotropic fluid pressure $P$ and the 4-vector fluid velocity $u_{\mu}$. The second term represents the purely Maxwell stress tensor, with $F_{\alpha\beta}$ being the usual Faraday tensor defined in terms of the magnetic vector potential $A_{\alpha}$ as $F_{\alpha\beta}=\partial_{\alpha}A_{\beta} - \partial_{\beta}A_{\alpha}$. 

According to \cite{Bonazzola:1993zz, Bocquet:1995je, gourgoulhon20123+}, the metric tensor can be expressed in spherical like coordinates $(r, \theta, \phi)$ within the 3+1 formalism as:
\begin{align}
ds^{2} = &-N^{2} dt^{2} + \Psi^{2} r^{2} \sin^{2}\theta (d\phi - N^{\phi}dt)^{2} \nonumber \\
 &+ \lambda^{2}(dr^{2} + r^{2}d\theta^{2}), 
\label{line}
\end{align}
with N, $N^{\phi}$, $\Psi$ and $\lambda$ being functions of the coordinates $(r, \theta)$. 
The equation of stationary motion ($\nabla_{\mu}T^{\mu\nu}= 0$) for perfect fluid with pure poloidal field can be expressed as \citep{Bocquet:1995je}:
\be
H \left(r, \theta \right) + {\rm{ln}}N\left(r, \theta \right)  + M \left(r, \theta \right) = const,
\label{equationofmotion}
\ee
where $H(r,\theta)$ is the logarithm of the dimensionless relativistic enthalpy per baryon and $M(r,\theta)$ the magnetic potential, which determines the magnetic field configuration:
\be
M \left(r, \theta \right) = M \left( A_{\phi} \left(r, \theta \right) \right): = - \int^{0}_{A_{\phi}\left(r, \theta \right)} f\left(x\right) \mathrm{d}x,
\ee
with a current function $f(x)$ as defined in \citep{Bonazzola:1993zz}. The Lorenz force induced by the magnetic field is proportional to $-\nabla M(r,\theta)$.  Magnetic stellar configurations are determined by choosing a constant current function $f_{0}$. The magnetic field strength in the star increases proportionally to $f_{0}$. Moreover, the macroscopic electric current scale with $f_{0}$ as $j^{\phi}= (\mathcal{E}+p)f_{0}$. Note that, in \cite{Bocquet:1995je} other possibilities for $f(x)$ were discussed, but the general conclusions remain the same.  The equations described above are solved numerically for different values of $f_0$ leading to neutron stars with different magnetic fields and thus different structures.

\section{Crust thickness of strongly magnetized stars}
 We describe the inner crust with the Skyrme (Sky) EoS, which is based on the effective nuclear interaction SLy of the Skyrme type. For more details on the composition and EOS calculation see \cite{douchin2001unified}. The structure of the inner crust, and its EoS, was taken from Baym, Pethick, Sutherland (BPS), based upon the Reid potential \citep{baym1971ground}. To describe the matter in the neutron star interior in the T = 0 approximation, we choose the APR EoS for the core \citep{akmal1998equation}, which is composed of protons, neutrons, electrons and muons. 

The equilibrium state of magnetized objects was discussed many years ago in\cite{ferraro1954equilibrium,chandrasekhar1956equilibrium}. More recently, many authors shown that the stellar radius increases due to magnetic fields, where the star expands in the equatorial direction and contracts at the pole \citep{mallick2014deformation,cardall2001effects,Bonazzola:1993zz,Bocquet:1995je,Chatterjee:2014qsa,Franzon:2015sya}. 
%In this work, we construct magnetized stellar models as described in section II, with a spherical-polar like coordinate system  specified by three numbers: the radial distance  $r$ of a point from the origin, its polar angle $\theta$, and the azimuth angle $\phi$. 

As already discussed in \cite{cardall2001effects}, the Lorentz force induced by magnetic fields reverses direction on the equatorial plane $\theta=\pi/2$ of the star. However, the consequence of such reversion on the outer layers of stars was not addressed in \cite{cardall2001effects}. Inspecting the outer layers, one observes that the crust can decrease or increase in size depending on the polar angle $\theta$, while the core expands in all directions. 

The crust thickness is defined as the difference between the stellar surface radius and the radius at the base of the crust where the crust-core transition takes place.  As already calculated in \cite{grill2014equation, xu2009nuclear}, the symmetry energy affects the size of the inner crust considerably. In addition, in \cite{fortin2016neutron} the importance of a consistent  matching between  the core  and  the crust regions was shown.  Although a more thorough study along this line would certainly be of interest,  for the purposes of our studies it is sufficent to use the   the Sly and BPS results,  in which the baryon number density at the crust-core transition is 0.076$ \,\rm{fm^{-3}}$. 

\begin{figure}
\includegraphics[height=8cm, angle=-90]{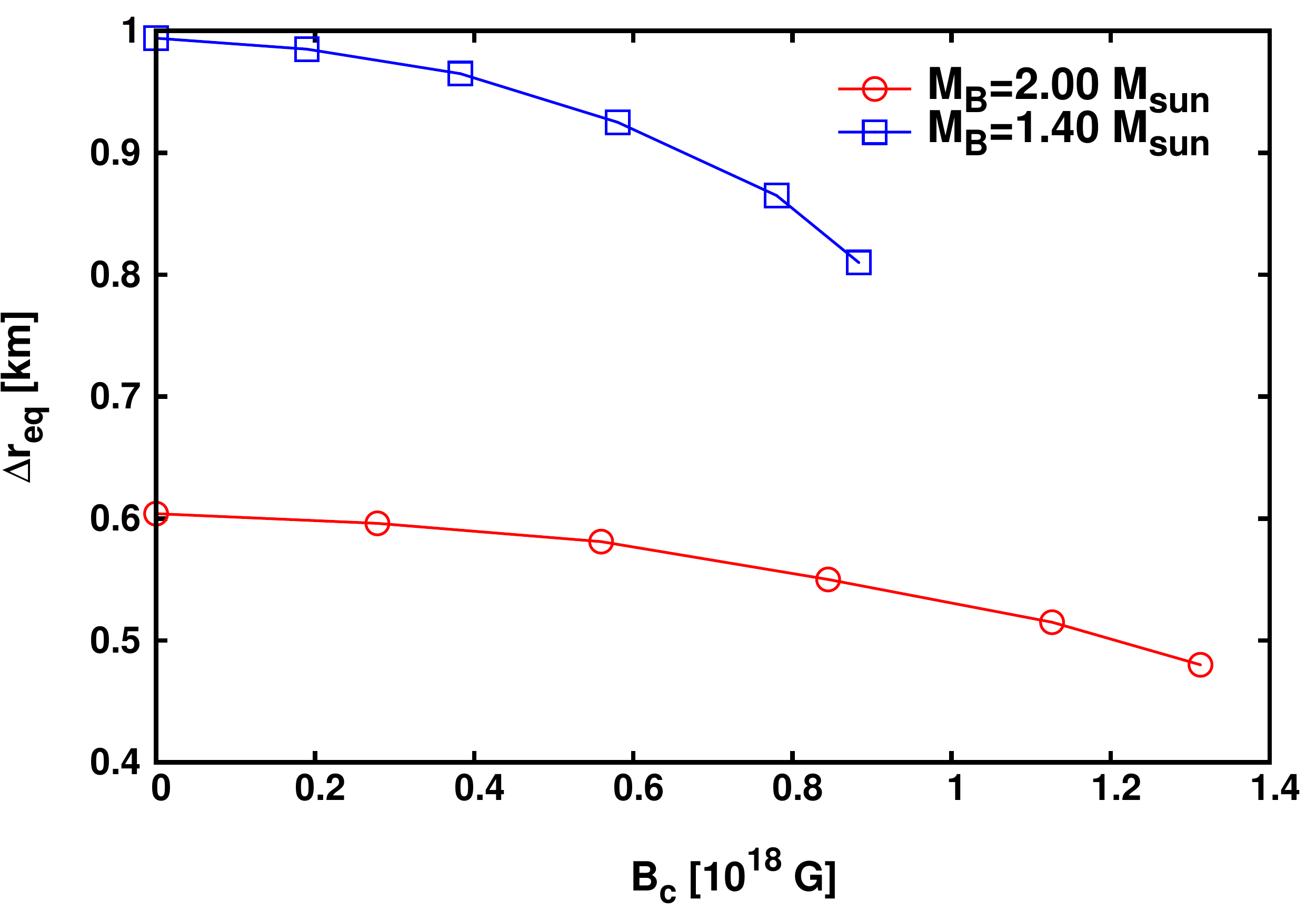}
\caption{Thickness of the crust $\Delta r_{eq}$ in the equatorial plane ($\theta=\pi/2$) as a function of central magnetic fields for stars at fixed baryon masses of $M_{B}=1.40\,M_{\odot}$ and $M_{B}=2.00\,M_{\odot}$.   }
\label{thickness_bfield}
\end{figure}

In order to illustrate the effects of strong magnetic fields on the neutron-star crust thickness, we shown in Fig.~\ref{thickness_bfield} the crust thickness in the equatorial plane ($\theta=\pi/2$), $\Delta r_{eq}$, as function of the central magnetic field, $B_{c}$, for stars at fixed baryon masses of $M_{B}=1.40\,M_{\odot}$ and $M_{B}=2.00\,M_{\odot}$, respectively. According to Fig.~\ref{thickness_bfield}, the maximum magnetic field reached at the center of the star $M_{B}=2.00\,M_{\odot}$ is 1.3$\times 10^{18}$ G, while a star with $M_{B}=1.40\,M_{\odot}$ has a central magnetic field of 0.9$\times 10^{18}$ G. As one can see in Fig.~\ref{thickness_bfield}, the crust thickness in the equatorial plane always reduces with the magnetic field.  Note that, stars with lower masses have a higher crust deformation. This is due to the fact that these stars  have larger crusts sprawled over a larger radius, and thus are easier deformed through the magnetic fields. On the other hand, the lower the stellar mass the lower the magnetic field at the stellar center. This is because  in a perfectly conducting fluid, the magnetic field lines move with the fluid, i.e., the magnetic field lines are `frozen' into the plasma and, therefore, the magnetic field strength is proportional to the local mass density of the fluid.  %\citep{mestel2012stellar}.

\begin{figure}
\includegraphics[height=8cm, angle=-90]{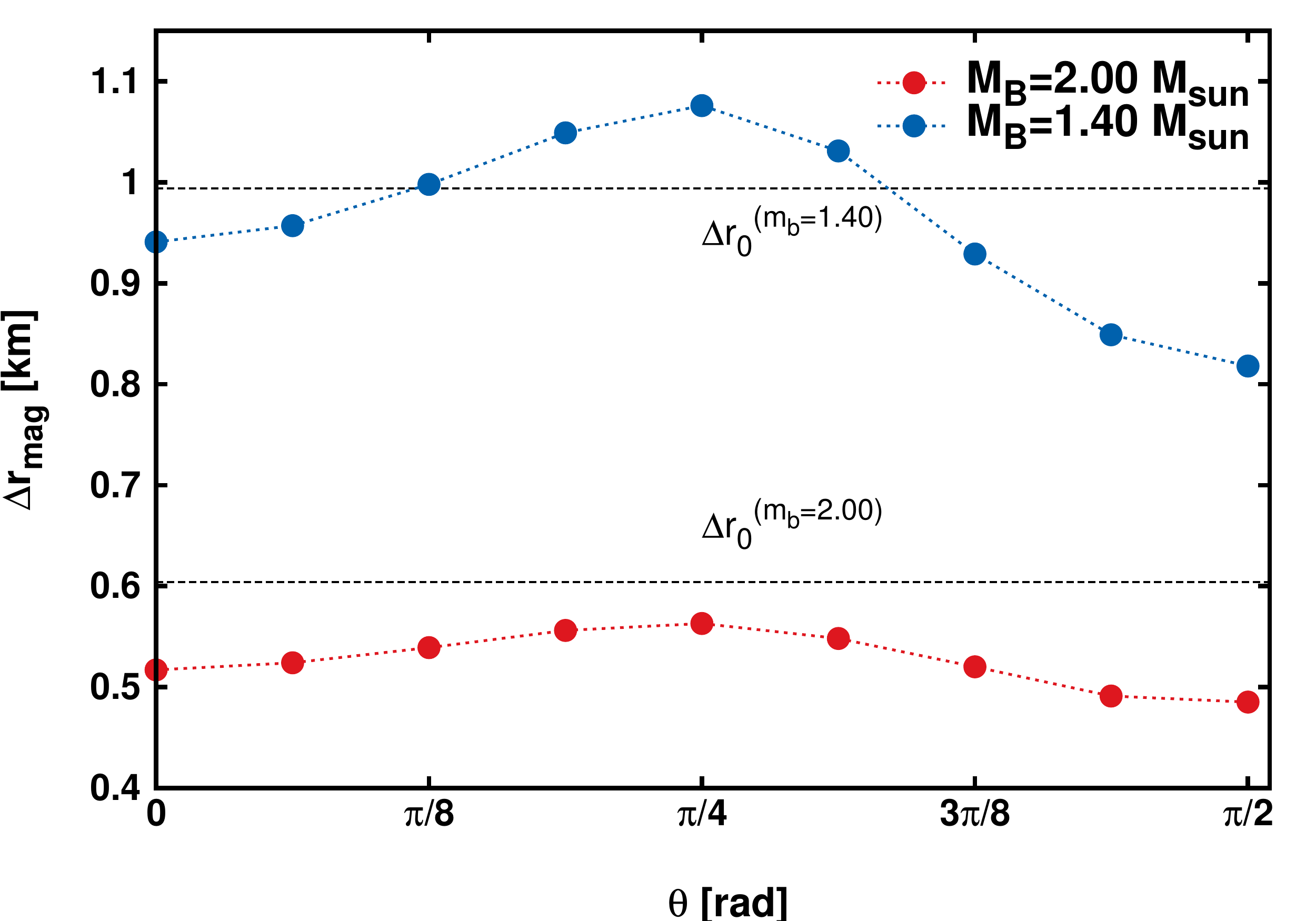} 
\caption{Crust thickness $\Delta r_{mag}$  in different angular directions $\theta$ for highly magnetized stars. These objects are the most magnetized stars as depicted in Fig.\ref{thickness_bfield}.  For at a fixed baryon mass of $M_{B}=1.40\,M_{\odot}$, the central magnetic field is $\sim 0.9\times 10^{18}$ G, while $M_{B}=2.00\,M_{\odot}$ has $\sim 1.3\times 10^{18}$ G.  The horizontal lines represent the crust thickness for spherical solutions (without magnetized fields), $\Delta r_{0}^{({1.40})}$=0.994 km and $\Delta r_{0}^{({2.00})}$=0.604 km for a star with $M_{B}=1.40\,M_{\odot}$ and $M_{B}=2.00\,M_{\odot}$, respectively.}
\label{thickness_angle}
\end{figure}

In order to study the effects of the Lorenz force on the crust of neutron stars, we show in Fig.~\ref{thickness_angle} the crust thickness as a function of the polar angle $\theta$ for the most magnetized stars obtained in Fig.~\ref{thickness_bfield}. The horizontal lines in Fig.~\ref{thickness_angle} correspond to the crust thickness for stars with baryon masses of $M_{B}=1.40\,M_{\odot}$ and $M_{B}=2.00\,M_{\odot}$, but without magnetic fields. In this case, the values for the crust thickness are $\Delta r_{0}^{({1.40})}$=0.994 km and $\Delta r_{0}^{({2.00})}$=0.604 km, respectively.

In Fig.~\ref{thickness_angle}, we depict stars that are deformed due to magnetic fields. As it was already calculated in \cite{cardall2001effects,Bocquet:1995je}, the stellar configurations can strongly deviate from spherical symmetry due to the anisotropy of the energy-momentum tensor in presence of strong magnetic fields. According to Fig.~\ref{thickness_angle}, the crust has a maximum expansion at $\theta=\pi/4$. For the less massive star ($M_{B}=1.40\,M_{\odot}$), the crust expands and becomes larger than its non-magnetized counterpart. From this point, the crust thickness reduces.  At the pole ($\theta=0$), the Lorenz force is zero by symmetry (no electric current at the symmetry axis), but the crust thickness is smaller than in the non-magnetized case.  This is a geometric effect due to the expansion of the star on the equatorial plane. On the other hand, the increase of the crust thickness followed by a reduction at diffrent polar angles is caused by the inversion of the direction of the Lorentz force inside the star.

\begin{figure}
\includegraphics[height=8cm, angle=-90]{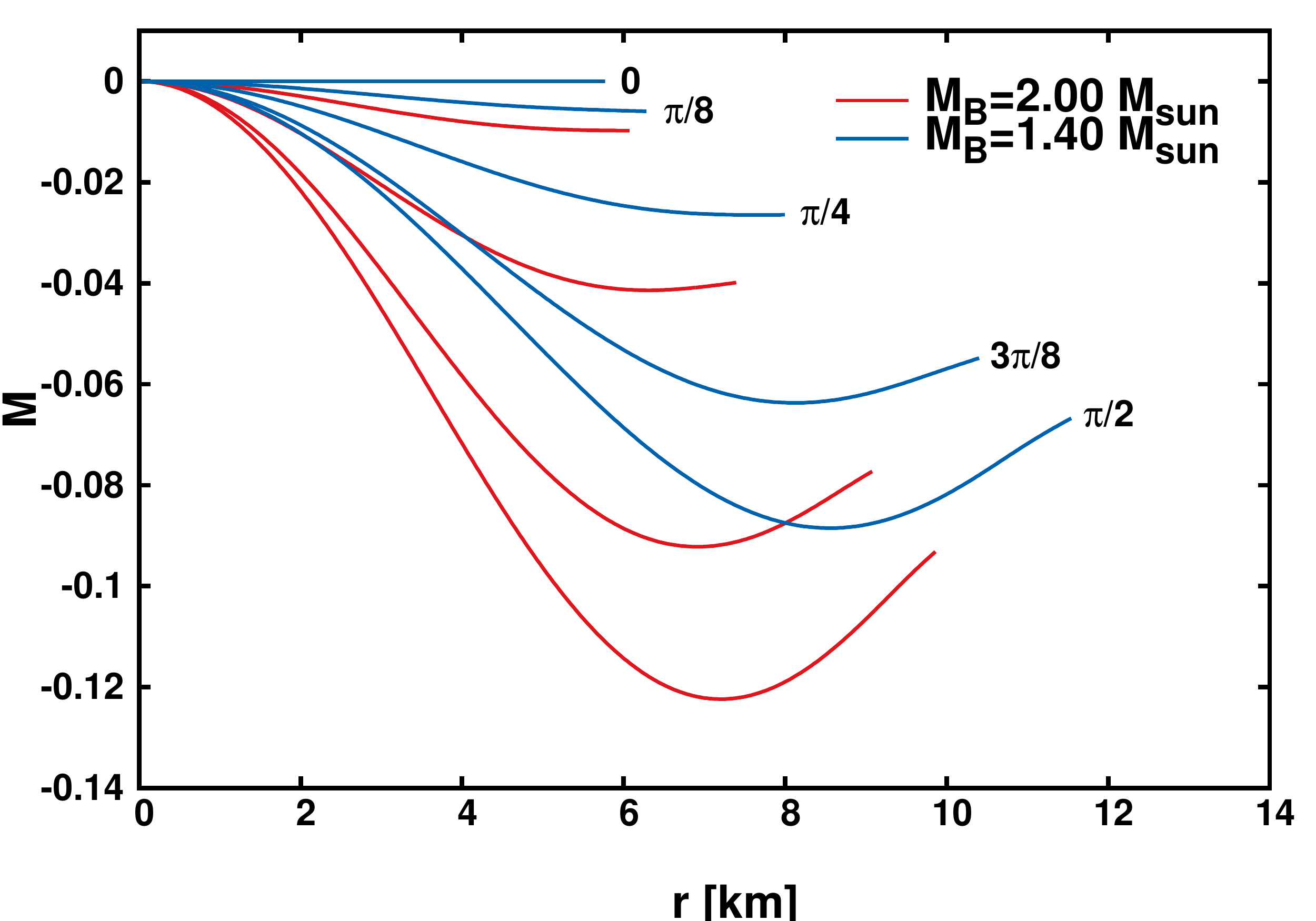}
\caption{Magnetic potential $M$ as a function of the coordinate radius in different directions inside the stars. The corresponding stars are depicted in Fig.~\ref{thickness_angle}.}
\label{magpotentical}
\end{figure}

In other to show the change of the Lorentz force according to its angular and radius distributions, we calculate in Fig.~\ref{magpotentical} the magnetic potential $ M(r, \theta)$ as a function of the coordinate radius and at different polar angle directions for the same stars shown in Fig.~\ref{thickness_angle}. As a result, one sees from Fig.~\ref{magpotentical} that the magnetic potential presents a minimum at higher angles, for example, at $\theta=3\pi/8$ and $\theta=\pi/2$ (for the star with $M_{B}=1.40\,M_{\odot}$). These values correspond to angles for which the Lorentz force reverses sign and, as a consequence, changes its direction in the star. At lower polar angles, $M(r, \theta)$  decreases monotonically and, therefore, the Lorenz force increases throughout the star which leads to an expansion both of the inner and the outer layers of the star. 

\begin{figure}
\includegraphics[height=8cm, angle=-90]{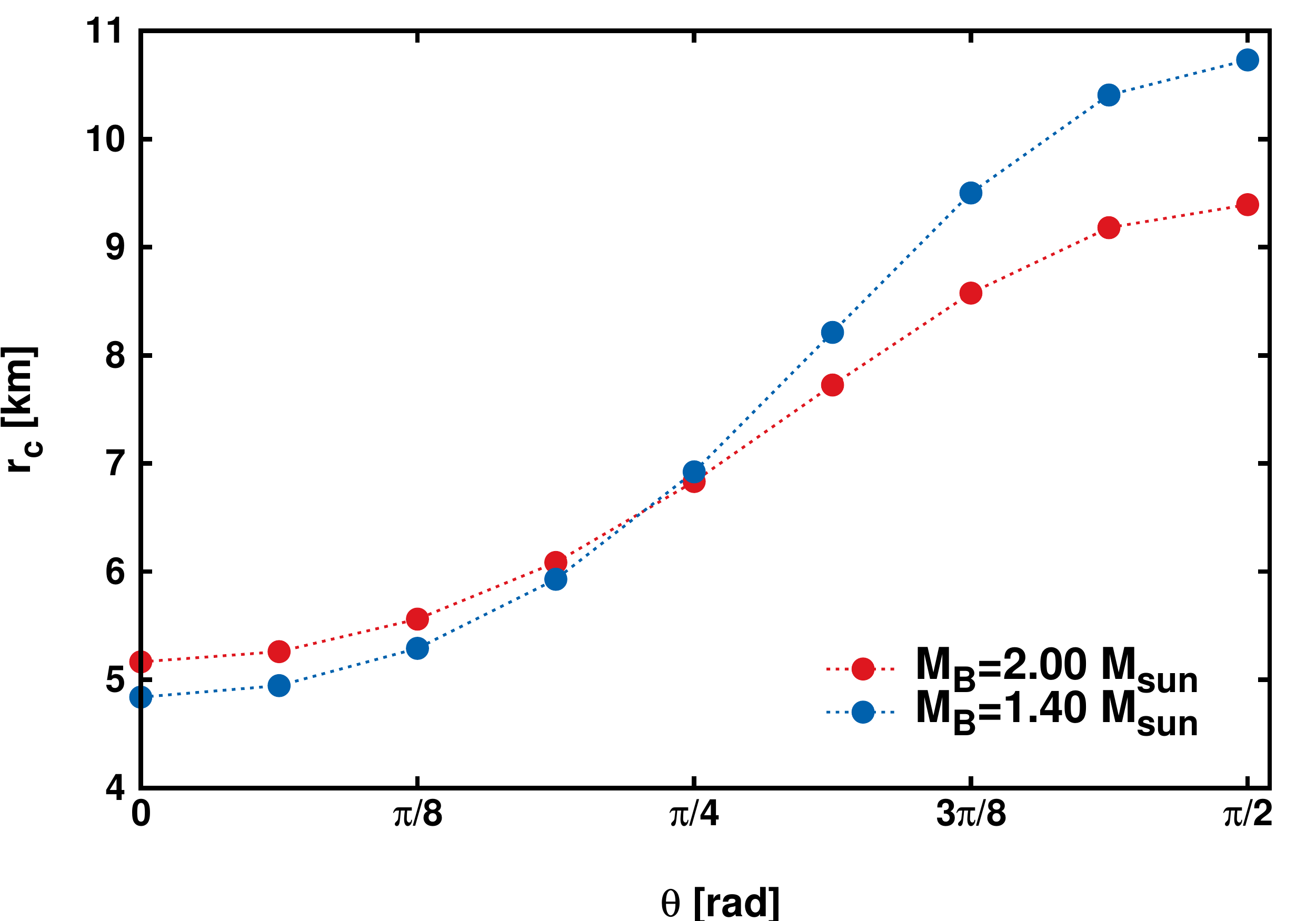} 
\caption{Core thickness for the same stars as shown in Fig.~\ref{thickness_angle}.}
\label{thickness_angle_core}
\end{figure}

We show in Fig.~\ref{thickness_angle_core} the size of the core for the same stars as those shown in Fig.~\ref{thickness_angle} and Fig.~\ref{magpotentical}. The results in  Fig.~\ref{thickness_angle_core} indicate that highly magnetized stars expands their cores in all directions $\theta$. Note that the curves have a inflection point  at $\theta=\pi/4$, which corresponds to the angle where the Lorenz force reverses direction inside these stars. Differently from the crust, which the increase or the decrease of the thickness depends on the polar angle $\theta$, the core always increases in size. This is due to the fact that the Lorentz force always acts outwards in the stellar core.

\begin{figure}
\begin{center}
	\includegraphics[height=8.8cm, angle=-90]{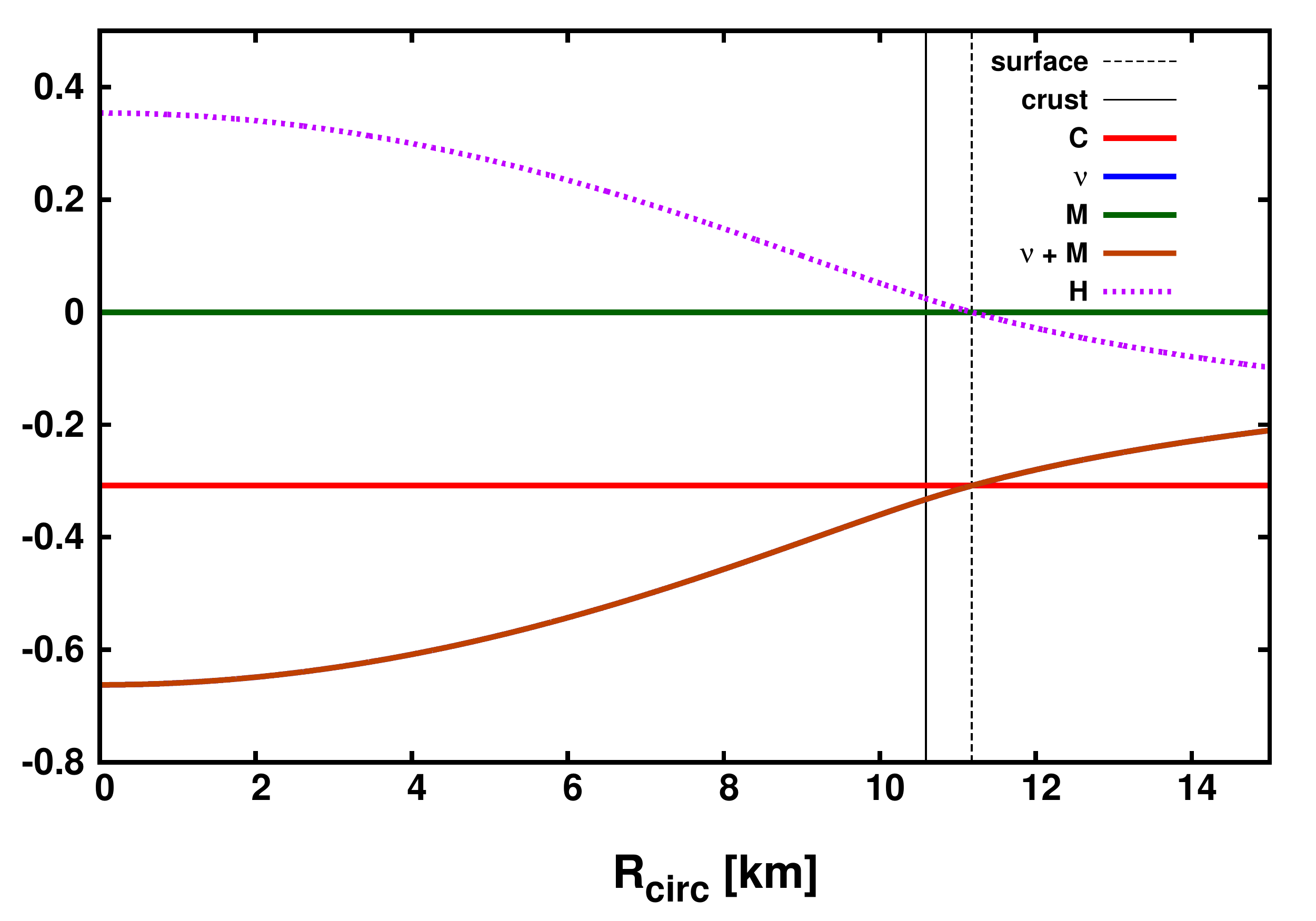} \quad
    \includegraphics[height=8.8cm, angle=-90]{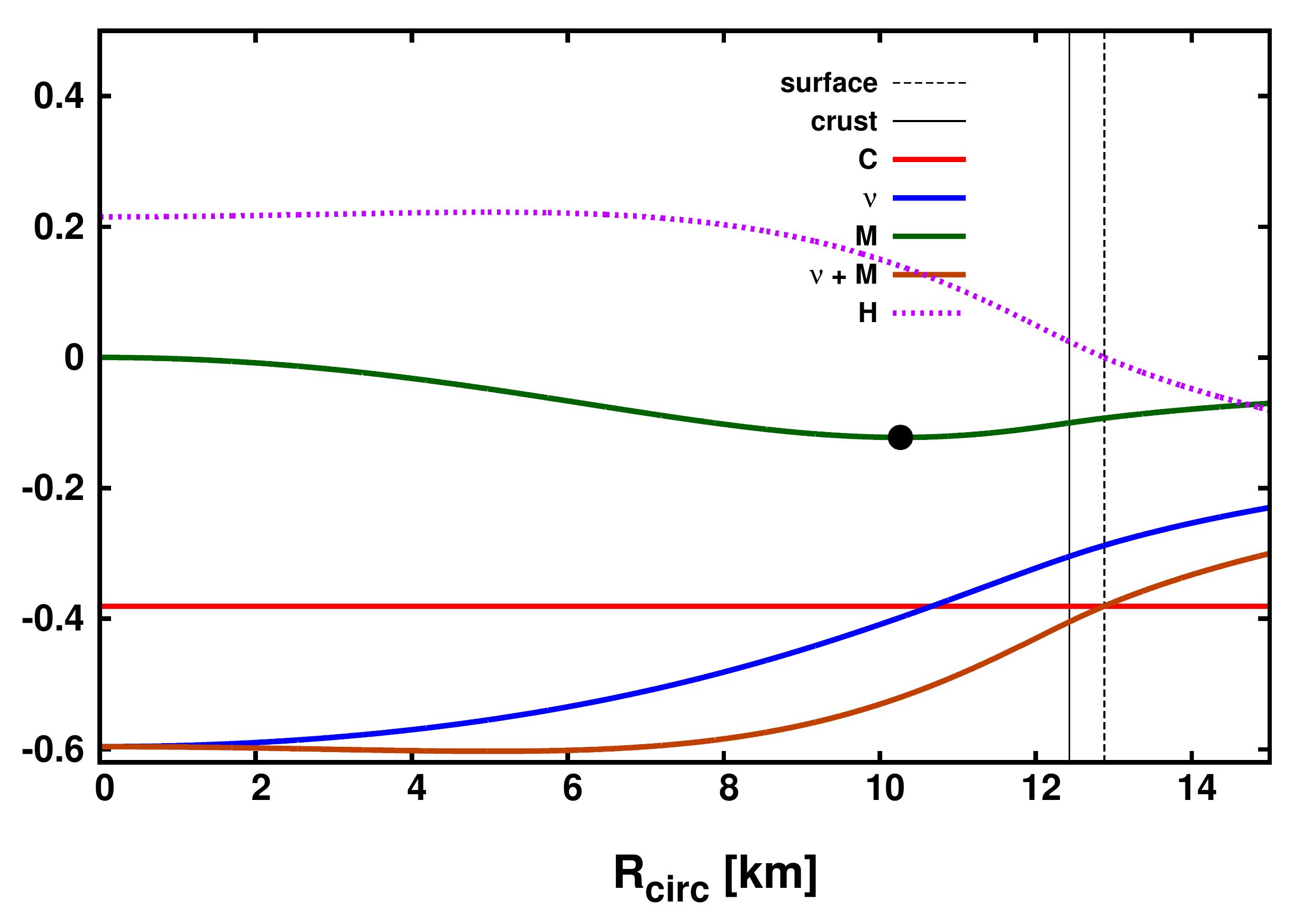}
    \includegraphics[height=8.8cm, angle=-90]{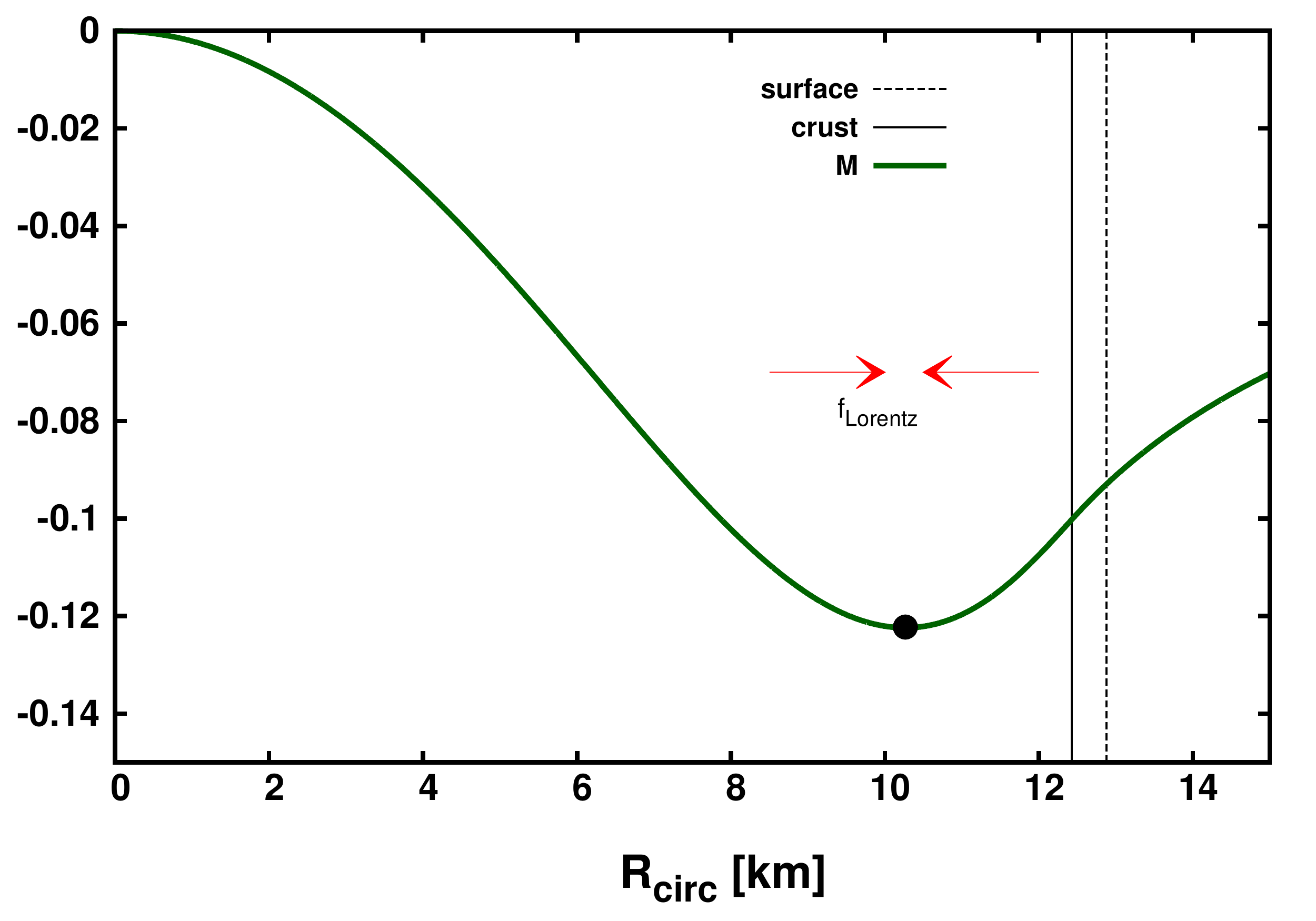}
\caption{Physical quantities presented in the equation of motion Eq.~\eqref{equationofmotion} as a function of the circular equatorial radius $R_{circ}$ for a star with $M_{B}=2.00\,M_{\odot}$. The upper plot represents a spherical and non-magnetized solution and the middle figure shows the same star, but highly magnetized. The central magnetic field is $1.3\times 10^{18}$ G. We highlight the direction of the Lorentz force in the star in the lower panel. The vertical lines are the crust line and the stellar surface. } \label{eqm}
\end{center}
\end{figure}

Fig.~\ref{eqm} depicts the physical quantities corresponding to the equation of motion in Eq.~\eqref{equationofmotion} as a function of the circular equatorial radius $R_{circ}$ for a star with $M_{B}=2.00\,M_{\odot}$.  $R_{circ}$ is defined as  $R_{circ}=\lambda(r_{eq},\pi/2)\,r_{eq}$, with $\lambda$ being the metric potential in Eq.~\eqref{line} and $r_{eq}$ the coordinate equatorial radius.  A detailed discussion about the coordinate system used in this work can be found in \cite{Bonazzola:1993zz}.

The upper plot in Fig.~\ref{eqm} represents a spherical and non-magnetized stellar solution. The central plot shows the quantities from Eq.~\eqref{equationofmotion}, i.e., $C, \nu, M$ and $H$, but taking into account magnetic fields. This is the same star as depicted in Fig.~\ref{thickness_angle} and  Fig.~\ref{thickness_angle_core}, respectively. In the bottom plot we highlight   the magnetic potential $M(r, \theta)$ and show the radii where the Lorentz force acts inwards and outwards inside the star. In all cases, the vertical lines represent the core-crust transition point and the stellar surface. As one can see, the star becomes bigger due to magnetic fields. However, the size of the crust decreases in the equatorial plane (see also Fig.~\ref{thickness_angle}).

For non-magnetized stars, the magnetic potential is $M(r,\theta)=0$. In addition, from Eq.~\ref{equationofmotion}, one has $H(r,\theta)+\nu(r,\theta) = const = C_{0}$. The constant $C_{0}$ can be calculated in every point in the star \citep{cardall2001effects}. Since the input to construct the stellar models are given at the stellar center, we choose $C_{0}=H(0,0)+\nu(0,0)$. 

In the upper plot in Fig.~\ref{eqm}, the surface of the star is found when the enthalpy goes to zero $H(r,\pi/2)=0$ and, therefore, from the equation of motion Eq.~\eqref{equationofmotion}, one gets $\nu=C_{0}$.  In the magnetized case (center plot), we see that the enthalpy reduces throughout the star and it reaches zero on the surface when $\nu + M = C_{0}$.  A similar analysis as those described above was already performed before for neutron stars in \cite{cardall2001effects} and for magnetized white dwarfs in \cite{Franzon:2015sya}.

The bottom plot in Fig.~\ref{eqm} shows the magnetic potential and the position and the direction (red arrows) of the Lorentz force inside the star. From the stellar center to the reversion point (black point), the Lorentz force acts outwards and, therefore, this part of the star expands. From the reversion to the crust-core radius (full black vertical line), the Lorentz force points inwards. This also holds true for the region between the crust-core transition and the stellar surface (dashed black vertical line).  As a result, the Lorentz force points always inwards in the crust region and, therefore, the crust reduces its size.  As a net effect, the star becomes bigger in the equatorial plane due to the increase of the core region.

\section{Thermal Relaxation of Magnetized Neutron Stars}
The study and observation of the thermal evolution of neutron stars has been established  as an important tool for probing the inner composition and structure of compact  stars \citep{Page2004,Page2006,Page2010,Negreiros2010a,Negreiros2012a,Negreiros2013b}. Many efforts have been dedicated towards a better comprehension of the thermal processes that may take place inside of neutron stars as well as the macroscopic structure effects that could affect the thermal evolution of such objects \citep{Negreiros2012,Negreiros2013b}.

Most thermal evolution calculations are performed under the assumption of spherical  symmetry and static structure composition, although efforts have been taken towards a self-consistent description of axis-symmetric neutron stars \citep{Aguilera2008,Negreiros2012} and objects with a dynamic structure evolution \citep{Negreiros2013b}. The work in \cite{Negreiros2012} shows that the thermal evolution 
of axis-symmetric neutron stars may be substantially different from that of spherically symmetric objects. Even though in the aforementioned paper the breaking of spherical symmetry is brought on due to rotation, it is reasonable to expect that a similar effect occurs if the spherical symmetry is broken due to the magnetic field as long as the resulting system also has an axis-symmetric structure. 

A particularly interesting result, discussed in \cite{Negreiros2012}, is the modification of the core-crust coupling time in axis-symmetric neutron stars. The core-crust coupling time is given by the duration it takes for the core and the crust of neutron stars to become isothermal. Due to the difference in composition between the core (comprised of hadrons and leptons, and possibly of deconfined quark matter \citep{Negreiros2012a})  and the crust (mostly heavy-ions in 
a crystalline structure and unbound neutrons in the inner crust) these two regions of the star have very distinct thermal properties, with substantially different neutrino emissivities, thermal conductivity and specific heat \citep{Page2006}. Due to such differences 
ordinarily the crust acts as a blanket, keeping the star's surface warm while the core cools down due to  stronger neutrino emission. Eventually the cold front, originated in the core arrives at the crust, cooling it off as it moves to the surface. At this moment a sudden drop 
in the stellar surface temperature is expected. Such drop signals the moment in which the neutron star interior (core and crust) is thermalized. The magnitude of the temperature drop will 
depend on whether or not fast cooling processes (mainly the Direct Urca process \citep{Page2004}) take place inside the neutron stars, as well as how pervasive superfluidity/superconductivity is in the core. The presence of fast cooling processes would lead to a deeper and sharper surface temperature drop, whereas the absence of fast processes (slow cooling) affects a smoother drop in surface temperature. The core-crust coupling time, also referred to as the cooling relaxation time, has been studied extensively in \cite{Gnedin2008}, in which the authors have found that the relaxation time, $t_w$, may be written as
\begin{equation}
t_w = \alpha t_1, 
\label{eq:tw}
\end{equation}
where $t_1$ is a characteristic time that depends solely on crustal microscopic properties such as thermal conductivity and heat capacity. It is also sensitive to neutron pairing, which may be present in the crust. It is important to notice, as pointed out in \cite{Gnedin2008}, the constant $t_1$ is almost independent of the neutron star model. This is reasonable since, regardless of the uncertainties with respect to the high density EoS, the composition of the crust is fairly known and understood. The constant $\alpha$ depends on stellar macroscopic properties and is given by 
\begin{equation}
\alpha = \left(\frac{\Delta R_C}{1\textrm{km}}\right)^2 \left( 1 - \frac{r_g}{R}\right)^{-3/2}
\label{eq:alpha}
\end{equation}
 with $\Delta R_C$ being the crust thickness, and $r_g = 2G M/c^2$ is the gravitational radius, with $M$ being the gravitational mass. 

In \cite{Gnedin2008} it  is found that the neutron star relaxation time ($t_w$) scales with the size of the crust according to eq.~(\ref{eq:tw}), more or less quickly, depending on how strong are the superfluid effects. Furthermore, it was also shown that the same conclusions hold for fast or slow cooling. 

Given the results put forth in \cite{Gnedin2008} and in \cite{Negreiros2012}, in addition to the results we show in this work regarding the crust properties of magnetized neutron stars, it is only natural to consider how the magnetic field, and the changes it brings about, would affect the relaxation time. For this reason, we follow the study of \cite{Gnedin2008} using the crust properties of magnetized neutron stars, as discussed in the sections above. It is important to notice that the study presented here  only  establishes an upper limit for the thermal relaxation time for magnetized neutron stars. The reason for this is that, whereas the results obtained in \cite{Gnedin2008} were for spherical symmetric stars, this is not the case for magnetized neutron stars, that have an axis-symmetric structure. In any the case, the change in the crust thickness should allow us to make a reasonably good estimate of the relaxation time of such objects. 
 
\begin{figure}
\includegraphics[height=8cm, angle=-90]{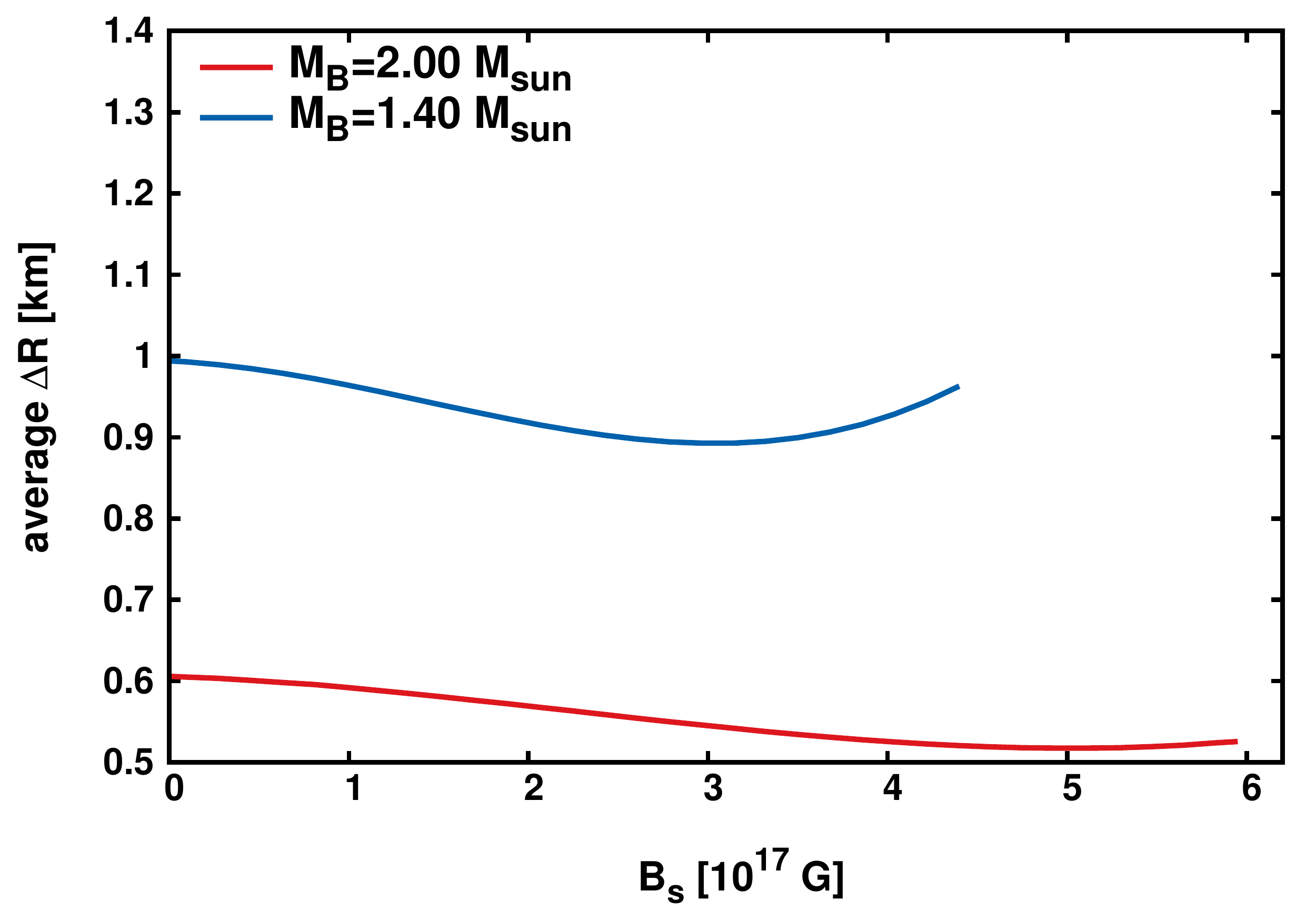} 
\caption{Average crust thickness for different values of surface magnetic fields.}
\label{AV_thick}
\end{figure}

In order to estimate how the modification of the crust properties will affect the the relaxation time, we consider in Fig.~\ref{AV_thick} the average crust thickness,  $\Delta R = \sum \Delta r (\theta)/N_{\theta}$, as a function of the stellar magnetic field. In other words, the average  $\Delta R$ is calculated for each value of the magnetic field, where $\Delta r (\theta)$ is the angular-dependent ($0\leq \theta \leq 2\pi$) crust thickness and $N_{\theta}$ the number of points in $\theta$.

As Fig.~\ref{AV_thick} shows, the crust becomes, on average, thinner for moderately high magnetic fields and thicker for the larger values of $B_s$. This is different than what happens in (also axially symmetric) rotating neutron star, whose crust always gets thicker with the increase of the rotational frequency \citep{Negreiros2012}. We believe that the reason for such difference is connected to the Lorentz force induced by the current distribution inside the star. For the magnetic fields studied in this paper the Lorentz force is attractive in the crust (same direction as the gravitational force), being stronger in the equatorial direction. This means that the crust will tend to become thinner on average for higher magnetic fields. On the other hand, one must note that the electromagnetic field in a general relativistic scenario has a dual role: it generates an electromagnetic force (in this case in the form of the Lorentz force, as just discussed) and with the electromagnetic energy it contributes to the curvature of space-time  (see \cite{Negreiros2009b,PicancoNegreiros2010a} and references therein for a more detailed discussion). Therefore, there are two competing effects: one tending to thin the crust and another causing it to be thicker (both on average). For moderately high magnetic field the former one is stronger, and the crust becomes thinner (on average), whereas the latter is dominant for very high magnetic field, causing the crust to become thicker (also on average). Evidently, the star with 1.40 solar mass, which has a lower gravity (i.e. curvature) is more susceptible to magnetic field effects and thus has a more pronounced effect, as illustrated in Fig.~\ref{AV_thick}. 

\begin{figure}
\includegraphics[height=8cm, angle=-90]{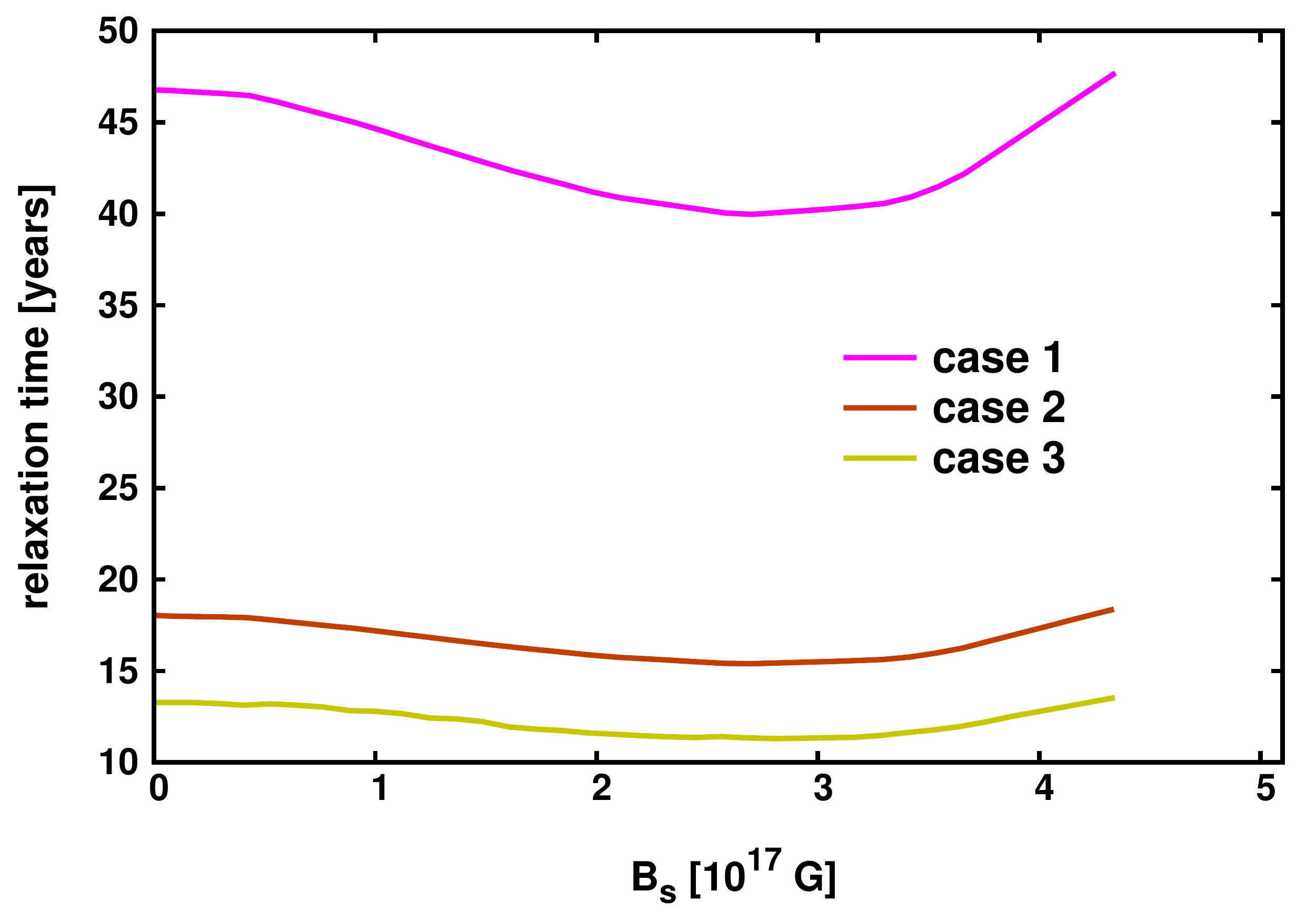} 
\caption{Relaxation time for the three cases studied of the 1.40 solar mass star.}
\label{RelT14}
\end{figure}

\begin{figure}
\includegraphics[height=8cm, angle=-90]{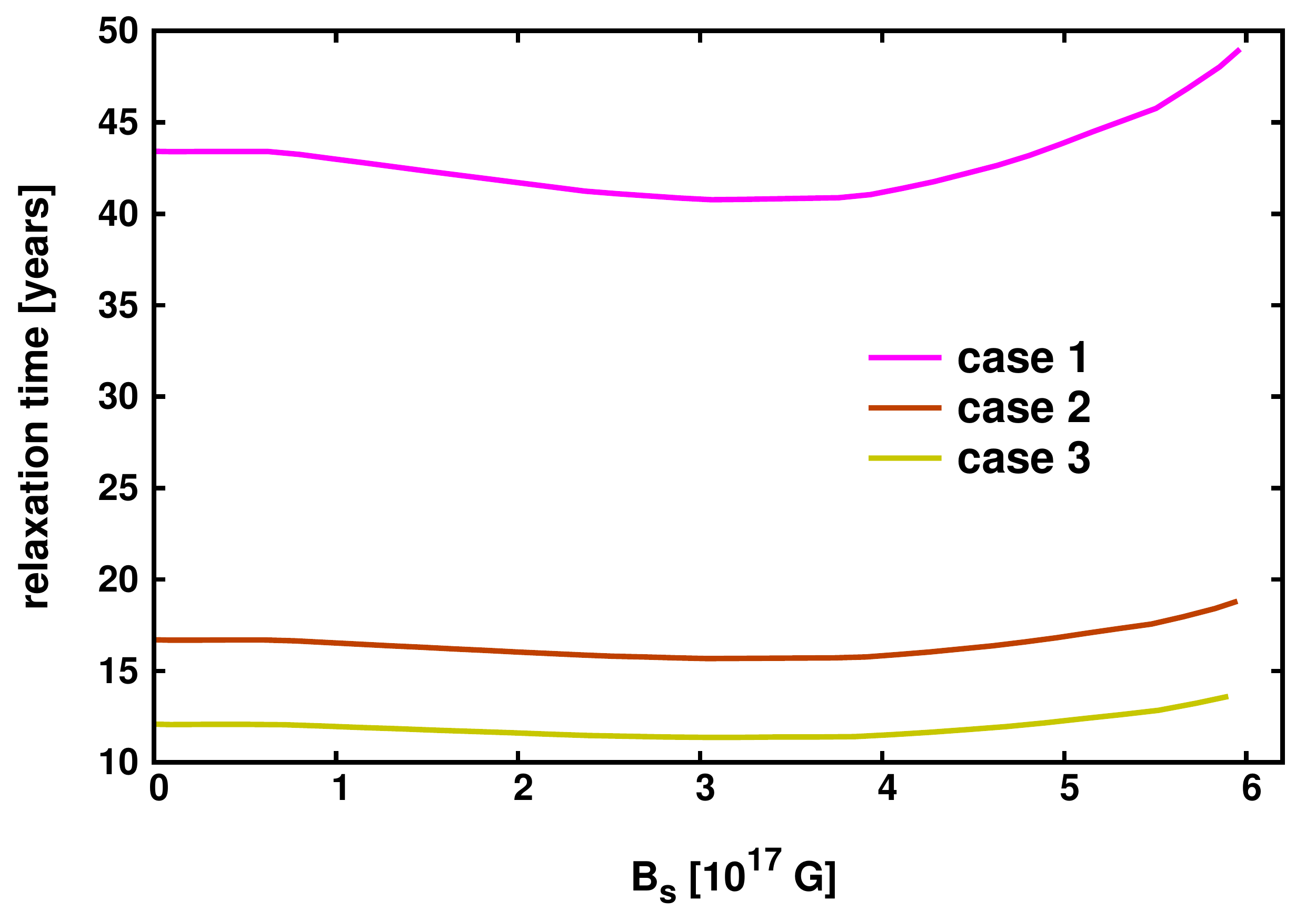} 
\caption{Relaxation time for the three cases studied of the 2.0 solar mass star.}
\label{RelT20}
\end{figure}

Following the steps of \cite{Gnedin2008} and using the average crust thickness, we now estimate the upper limit for the relaxation time of magnetized neutron stars. For that, as in \cite{Gnedin2008}, we consider three situations, identified by three 
different values for the normalization constant $t_1$, namely: $t_1 = 28.8$ years (case 1), 
associated with absence of superfluidity in the crust; $t_1=11.1$ years, associated with 
weak superfluidity in the crust (case 2); and $t_1 = 8.2$ years for the case of strong 
crustal superfluidity (case 3). The results are shown in Figs.~(\ref{RelT14}) and 
(\ref{RelT20}) for the 1.4 and 2.0 solar mass stars, respectively.

As expected due to the linear dependence on crust thickness, the relaxation time 
initially decreases as a function of the magnetic field and increases for higher values of $B$.
Once again, the 1.40 solar mass star is more susceptible to the effects of the magnetic field.

\begin{figure}
\includegraphics[height=8cm, angle=-90]{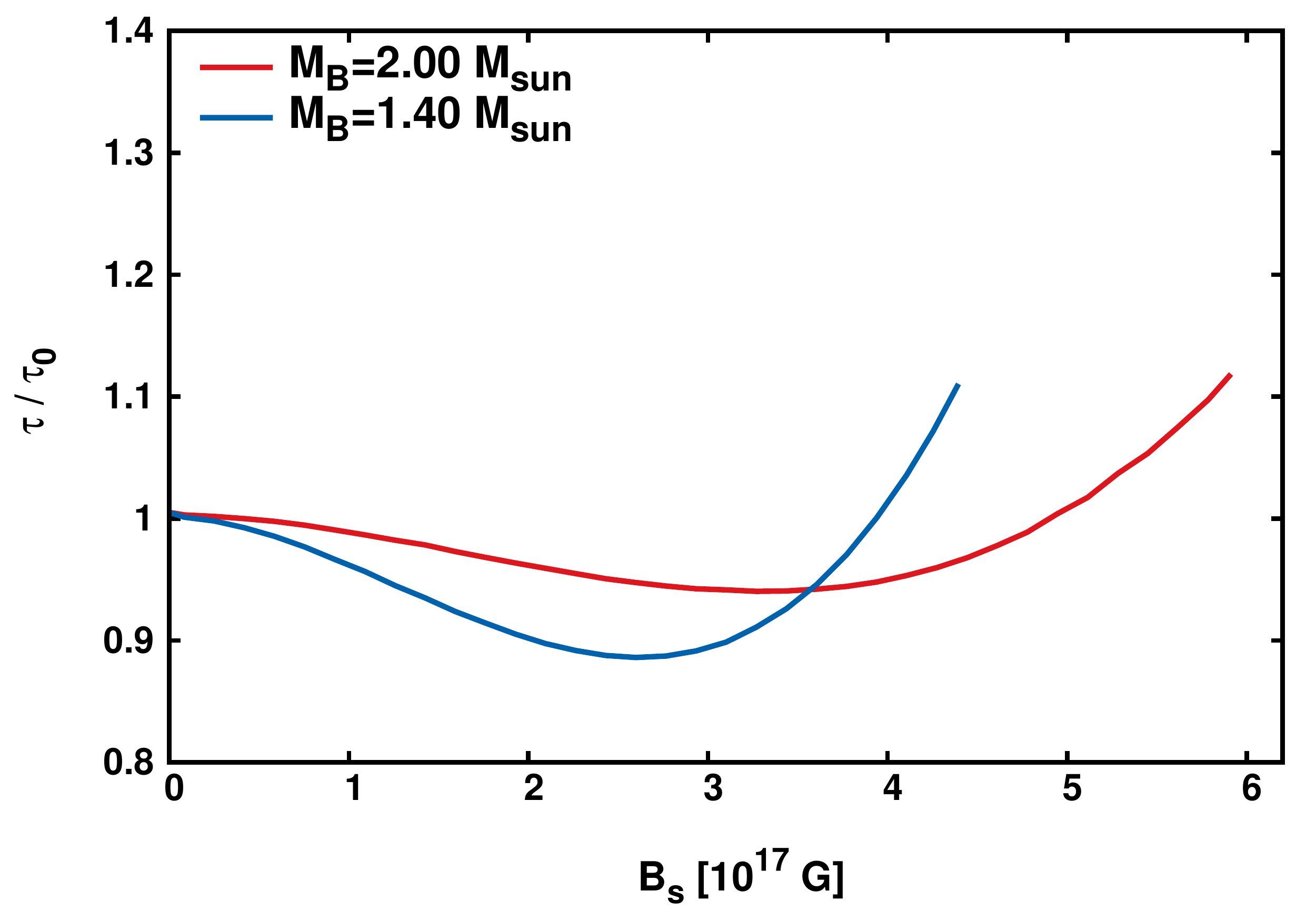} 
\caption{Relaxation time over the relaxation time at zero magnetic field as a function of surface magnetic fields for star at 1.4 and 2.0 solar masses.}
\label{norm_tau}
\end{figure}

If one wants to eliminate the uncertainties associated with the normalized time ($t_1$, which is connected to the extent and intensity of superfluidity in the crust) one can evaluate $\tau_w/\tau_0$, where $\tau_0$ is the relaxation time for the case with $B_s = 0$. This result is shown in Fig.~\ref{norm_tau}. 

\section{Conclusion}
 %Our calculations of magnetized neutron star  crust  can be applied also to shear mode oscillations of magnetars. In particular, shear modulus of the crust is sensitive not only to the to the crust compositions, but also the crust thickness, which can increase or decrease according to be the  to  We have observed that our model of the magnetised crust might
%explain the observed shear mode frequencies quite well.

%According to Ref.~\cite{nandi2011inner,PhysRevC.92.035802},  strong magnetic fields  modify the properties of the neutron star crust in different ways. First, the motion of charged particles is quantized in Landau levels, see e.g. \cite{landau1965quantum}. This modifies the electrical conductivity [citar efeito  de B na condutividade termica], which is important, for example, for the cooling process in neutron stars. Second,  the magnetic field  favors the appearance of heavier nuclei at low pressures, modifying the crust composition.  On the other hand, as shown in Refs.~\cite{Franzon:2016iai,Chatterjee:2014qsa}, the leading contribution to macroscopic properties of neutron stars, like mass and radius, do not depend significantly on the magnetic field dependent EOS or the inclusion of magnetization.  Although we do not expect a considerable change in our calculation while taking these effects into consideration for the crust, a  fuller understanding  of  this  process requires the study of such effects in order to have fully self-consistent stellar models. Calculations along this line are already in progress.
In this work, we studied strong poloidal magnetic fields effects on the global structure of the crust in stationary highly magnetized neutron stars. We self-consistently account for the Lorentz force, with current density bounded
within the star, by solving the coupled equilibrium equations for magnetic and gravitational fields consistently, taking into account the stellar deformation due to anisotropies induced by magnetic fields. We have employed typical and well-known equations of state to describe the inner and the outer crust of neutron stars.  

We found that that the size of the crust change according to its angular-dependent distribution inside the star. As already discussed  in \cite{cardall2001effects} and later on in \cite{Franzon:2015gda} in the context of magnetized white dwarfs, the magnetic force is zero along the symmetry axis and its  direction depends on the current distribution inside the star. Moreover, the magnetic field changes its direction and, therefore, the Lorenz force reverses the direction inside the star. In our case,  this can be seen from  the  change in behaviour of the magnetic potential $M(r, \theta)$ around $\rm{R_{circ}} \sim$ 10 km in Fig.~\ref{eqm}.

In this work, we have taken steps to estimate how the magnetic field, and the consequent modification of the crust and space-time of the neutron star may affect the thermal evolution of neutron stars. In addition to the expected effects that a magnetic field may have on the microscopic composition, we have shown that the change in crust geometry may be very relevant to the overall cooling of neutron 
stars. Using the crust average thickness as a parameter, we estimated the upper limit for the thermal evolution relaxation time, which is the time scale for the core-crust thermal coupling. We have found that the crust thickness (on average), as a function of the quantity responsible for the breaking of spherical symmetry (the magnetic field in this case) gets smaller before growing. This is substantially 
different from other axially symmetric neutron stars, such as rotating objects for instance. In the latter case the crust gets always thicker as function of the spherical symmetry breaking quantity (rotation/angular momentum in that case). We conclude that the reason for such behavior lies in the dual role of electromagnetic field in a general relativistic scenario, whose energy contribute to curvature in 
addition to the electromagnetic traditional interaction. We have found that for the stars studied, there are two competing effects, one is the Lorentz's force that tends to make the crust thinner, whereas the gravitational contribution of the magnetic field tends to make the crust thicker. For moderately high magnetic fields the former wins and the crust gets thinner on average, whereas for extreme values of $B$ the latter is dominant, making the crust thicker, overall. This behavior is reflected on our estimates of the core-crust coupling time, which, as a function of the surface magnetic field, gets  initially smaller, and increases for higher values of $B$. Such result is interesting, since one would be inclined to believe that the relaxation time would increase monotonically with $B$. One also must note that the overall geometry of the star becomes more oblate with the increase of $B$, so such assumption would be reasonable. However, due to the Lorentz's force acting on the crust, its size is reduced for moderate values of $B$. 

Our results represent magnetostatic equilibrium conditions. The stability of these equilibria are beyond the scope of this initial discussion on the possible observable through the crust geometry in highly magnetized stars. Note that, purely poloidal or purely toroidal magnetic field configurations undergo intrinsic instabilities related to their geometries \citep{markey1973adiabatic, tayler1973adiabatic, wright1973pinch, flowers1977evolution, lander2012there, braithwaite2006stability, ciolfi2013twisted, lasky2011hydromagnetic, marchant2011revisiting,  mitchell2015instability}. In this context, several calculations have also shown that stable equilibrium configurations are obtained for magnetic fields composed of both a poloidal and  a toroidal component \citep{armaza2015magnetic, prendergast1956equilibrium, braithwaite2004fossil, braithwaite2006stable, akgun2013stability}. In addition, we obtained surface magnetic fields values above those observed so far in neutron stars. Nevertheless, according to the Virial theorem, the magnetic fields reached at the center of neutron stars are expected to be so high as the magnetic field values found in this work. Although we have restricted to purely poloidal magnetic field, which is not the most general case, we have shown, in a fully general relativity way, that strong magnetic field affects significantly the crust geometry and its size. As a result, the thermal properties of these objects as the cooling relaxation time is affected correspondingly.

Evidently our calculations should be seen as an upper limit for the relaxation time, since full thermal evolution calculations such as in \cite{Negreiros2012} would be necessary. In any case, the interesting behavior of the crust geometry warrants further investigation and shows that the thermal behavior of magnetized neutron stars may not be straight forward. Studies in which the magnetic field changes over time may lead to even more interesting and unexpected behavior. Current efforts are being made towards the investigation of different current distributions (which may lead to the Lorentz's force having a different effect) as well as full 2D thermal evolution calculation of magnetized neutron stars.

%In summary, in this study we have modelled highly magnetized hybrid stars with a purely poloidal magnetic field  in a fully general relativity way. A poloidal field geometry is not the most general case, but by using these restrictive assumptions we have seen that the microphysics of a quark-hadron phase transition can potentially affect the macroscopic and observable magnetic fields of such stars.  Additionally, in order to better understand the stars presented here, the thermal evolution of such configurations will be reported in upcoming work. 

 %Our models of rapidly rotating protoneutron stars
%were based on several simplifications, which were neces-
%sary in order to make the problem tractable. In order
%to avoid difficulties and/or ambiguities of the largely un-
%known âreal situationâ, we restricted ourselves to study-
%ing idealized, limiting cases. In many places we introduced
%approximations, which were crucial for making numerical
%calculations feasible.

%In this work, we developed a self-consistent approach to
%determine the structure of neutron stars in strong mag-
%netic fields, relevant for the study of magnetars.
\section{Acknowledgements}

B. Franzon acknowledges support from CNPq/Brazil,
DAAD and HGS-HIRe for FAIR. R. Negreiros acknowledges the  
financial support from CAPES and CNPQ. S. Schramm acknowl-
edges support from the HIC for FAIR LOEWE program.
The authors wish to acknowledge the 'NewCompStar'
COST Action MP1304.

\bibliography{biblio} % if your bibtex file is called example.bib

% Don't change these lines
%\bsp	% typesetting comment
\label{lastpage}
\end{document}